\documentclass[twocolumn]{aastex62}
\usepackage{lineno}

\usepackage[caption=false]{subfig}
\usepackage{graphicx}
\graphicspath{ {images/} }
\usepackage{hyperref}
\usepackage{amsmath}
\usepackage{booktabs}
\usepackage{xcolor}
\usepackage[percent]{overpic}
\usepackage{euler}

\begin{document}

\title{Evidence of High-energy Neutrinos from SN1987A by Kamiokande-II and IMB}

\author{Yuichi Oyama}
\affiliation{High Energy Accelerator Research Organization (KEK), Tsukuba, Ibaraki 305-0801, Japan}
\affiliation{J-PARC center, Tokai, Ibaraki 319-1195, Japan}

\submitjournal{The Astrophysical Journal}
\received{October 24, 2021}\revised{December 7, 2021}\accepted{December 11, 2021}\published{February 3, 2022}

\begin{abstract}
High-energy neutrinos from SN1987A were searched for using upward-going muons
recorded by the Kamiokande-II experiment and the IMB experiment. 
Between August 11 and October 20, 1987, and from an angular window of
$10^{\circ}$ radius, two upward-going muon events were recorded by Kamiokande-II,
and also two events were recorded by IMB.
The probability that these upward-going muons
were explained by a chance coincidence of atmospheric neutrinos was calculated
to be 0.27\%. This shows a possible evidence of high-energy neutrinos
from SN1987A.
\end{abstract}

\section{Introduction}
Supernovae are considered as a primary source of
ultrahigh-energy cosmic-ray protons and heavier
ions \citep{Gunn,Goldreich,Scott,Colgate,Blandfold}.
High-energy ($>$ GeV) neutrinos can be generated from
supernova remnants immediately after the explosion.
In this scenario, protons accelerated by newly born
neutron stars collide with a sufficiently thick gas
that initially comprised the envelope of the progenitor
and subsequently diffused due to supernova explosion.
Charged pions and kaons produced in the collisions decayed
into high-energy neutrinos.
Observations of high-energy neutrinos from a supernova remnant will confirm
the existence of an acceleration mechanism of ultrahigh-energy protons \citep{Berezinsky,Sato1}.

\begin{table*}
\caption{
A list of upward-going muon events from the direction of SN1987A
detected by Kamiokande-II and IMB
The date and time of the events, their celestial positions, angular deviation from SN1987A,
and the event categories are listed. ``thru" indicates upward through going muons and
``stop'' indicates upward stopping muons.
In the last column, the events from the $10.0^{\circ}$ angular window
between August 11 and October 20, 1987 are indicated by open circles.
}
\begin{center}
\begin{tabular}{cccrclc}
\hline
\hline
  &  ~~~Date/Time (UT) & (RA,$\delta$)~~~& $\Delta\theta_{\rm SN}$&~~~~$\cos\Delta\theta_{\rm SN}$ &~~category&~~selected\\
\hline
1~~ & ~~~~1987-08-12~~05:08:27~~~~& (97.1$^{\circ}$,$-$67.5$^{\circ}$)~~~ & 5.2$^{\circ}$ &~~0.996~~~&~~IMB-thru&~~$\circ$~\\
2~~ & ~~~~1987-09-19~~11:39:10~~~~& (67.2$^{\circ}$,$-$67.9$^{\circ}$)~~~ & 6.2$^{\circ}$ &~~0.994~~~&~~Kam-thru&~~$\circ$~\\
3~~ & ~~~~1987-10-16~~10:17:49~~~~& (60.0$^{\circ}$,$-$73.9$^{\circ}$)~~~ & 8.7$^{\circ}$ &~~0.988~~~&~~IMB-thru&~~$\circ$~\\
4~~ & ~~~~1987-10-18~~16:16:47~~~~& (18.8$^{\circ}$,$-$63.5$^{\circ}$)~~~ & 25.4$^{\circ}$ &~~0.903~~~&~~Kam-stop&\\
5~~ & ~~~~1987-10-18~~18:15:00~~~~& (95.9$^{\circ}$,$-$61.2$^{\circ}$)~~~ & 9.5$^{\circ}$  &~~0.987~~~&~~Kam-stop&~~$\circ$~\\
6~~ & ~~~~1988-01-13~~03:23:27~~~~& (67.4$^{\circ}$,$-$64.0$^{\circ}$)~~~ & 8.4$^{\circ}$  &~~0.989~~~&~~IMB-stop&\\
\hline
\hline
\end{tabular}
\end{center}
\label{tab:eventlist}
\end{table*}

\begin{figure*}
\centering
\includegraphics[width=30pc]{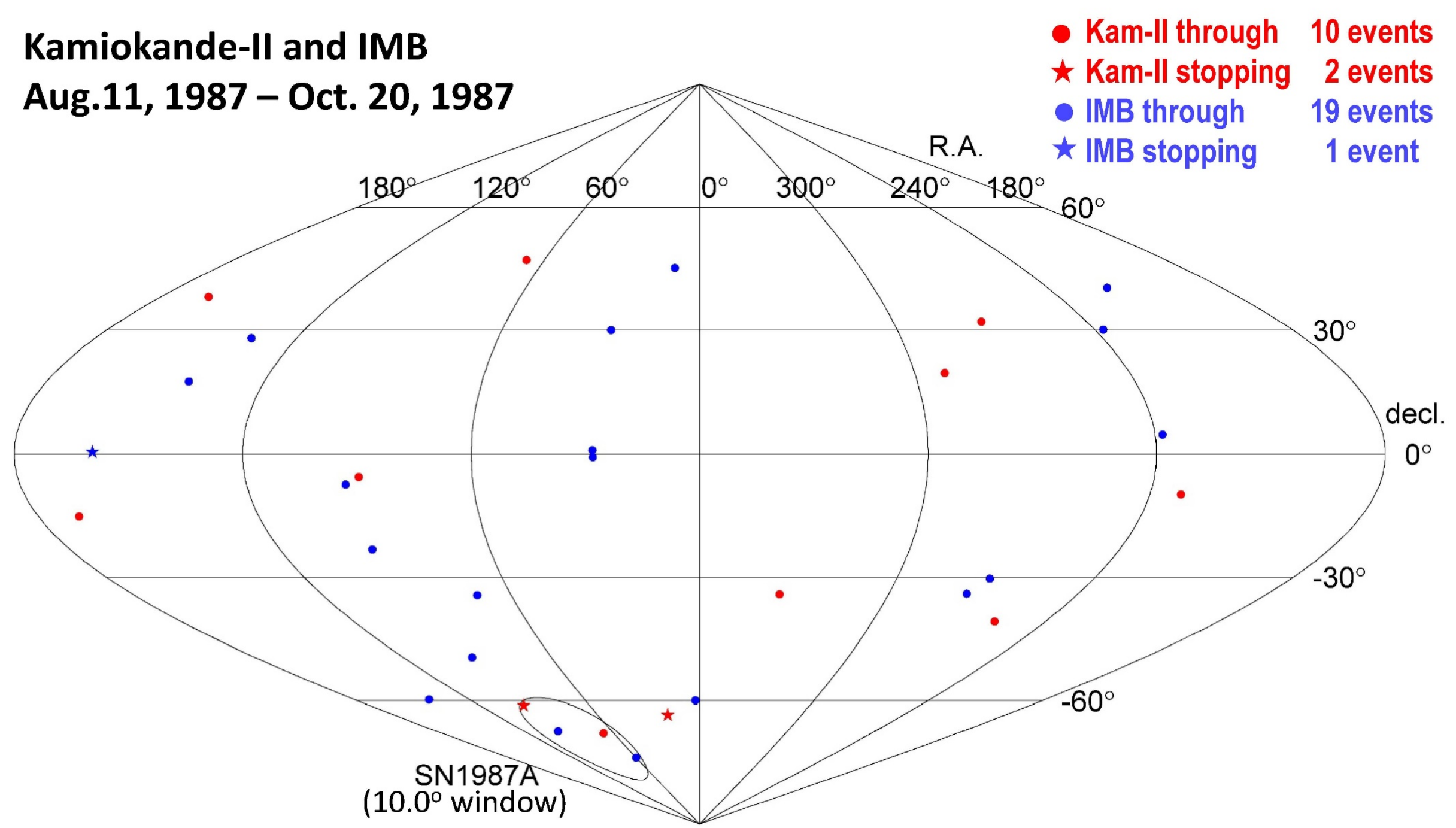}
\caption{Celestial map of upward-going muon events observed by Kamiokande-II
and IMB between August 11 and October 20, 1987.
The $10.0^{\circ}$ angular window around SN1987A is also shown.
}
\label{fig:skymap}
\end{figure*}

The supernova SN1987A (RA=83.9$^{\circ}$, $\delta$=$-$69.3$^{\circ}$) \citep{SN1987A,SN1987Afull,IMB} 
provided an excellent opportunity
to search for such processes experimentally.
High-energy neutrinos can be observed as upward-going muons in underground detectors.
The expected number of upward-going muon events from SN1987A
and their time structures were previously reported \citep{Gaisser,Sato2}.
Although they are model-dependent,
10$\sim$100 upward-going muon events could be observed
by $\sim$100~m$^{2}$ detectors within one year from the explosion.

\section{The data}
After the SN1987A explosion, underground detectors, including
Kamiokande-II and IMB, searched for upward-going muon events from SN1987A.
The results from Kamiokande-II were published until
191 days after the explosion \citep{SN1987Aupmu}.
The results until 385 days from the explosion were also reported \citep{Oyama}.
One upward through going muon event
was detected on the 209th day (September 19, 1987)
within a $7.0^{\circ}$ angular window.
However, it was not statistically significant because the
expected atmospheric neutrino background was approximately 0.4 \citep{Oyama}.

In addition, the Kamiokande-II experiment observed 2 upward stopping
muon events on October 18.
At that time, a systematic analysis of upward stopping muons
was not performed owing to their small event number and poor
angular correlations with parent neutrinos.
However, some upward stopping muons were found during
a visual scan in the upward through going muon selection because of
a similarity of the event topologies. 
The results were not published officially because the collaboration
concluded that
even if upward stopping muons are combined with the through-going
muon candidate detected on September 19, the statistical significance
is still poor to claim as evidence.

The IMB experiment accumulated upward-going muon events
throughout the entire experimental period.
A total of 666 through-going events between September 1982 and December 1990,
and 34 stopping events between May 1986 and February 1990 were recorded.
Upward-going muon events from the direction of SN1987A were examined,
and three possible candidates were found.
However, they also concluded that the statistical significance
was not satisfactory for a publication.

Recently, it was ascertained that two experiments both independently 
obtained statistically insufficient results, of which neither were published.
However, the combined results seemed more publication worthy.
Therefore, in this study, the results of the combined analysis are presented.

Details of the analysis can be found in the original
papers \citep{SN1987Aupmu,Kamupmu,IMBupmu} and are not discussed here.
A list of three possible candidates from each experiment are given in Table~\ref{tab:eventlist}.

\begin{table*}[!t] 
\caption{
Number of upward-going muon events detected by Kamiokande-II
and IMB. The first column is the event category. The second column indicates
number of candidates in each category. The third column shows the expected atmospheric
neutrino background
from the time/angular window. The calculation and the results of
the probability $N \geq N_{cand}$ are shown in the final two columns.
During calculation, the Poisson distribution function
${\rm Poi}(N|\lambda)={{e^{-\lambda} \lambda^{N}}\over{N!}}$
was used.
}
\begin{center}
\begin{tabular}{lllll}
\hline
\hline
Event   ~~~~          & Number of  &~~~~Expected &~~~~Calculation& Probability of\\
Category~~~~          & candidates &~~~~atmospheric $\nu$  &  & $N \geq N_{cand}$\\
                            & ($N_{cand}$) &~~~~background ($\lambda$) &  & \\
\hline
Kam-thru &~~~~~1&~~~~~~~0.127&~~~1$-$Poi(0$|$0.127)&~~~0.119\\ 
Kam-stop &~~~~~1&~~~~~~~0.026&~~~1$-$Poi(0$|$0.026)&~~~0.026\\ 
IMB-thru  &~~~~~2&~~~~~~~0.315&~~~1$-$Poi(0$|$0.315)$-$Poi(1$|$0.315)&~~~0.040\\ 
IMB-stop &~~~~~0&~~~~~~~0.029&~~~~~--- &~~~--- \\
\hline
\hline
\end{tabular}
\end{center}
\label{tab:probability}
\end{table*}

\section{The statistical significance and discussion}
For statistical significance calculations, the time and angular
window were assumed as follows.
\begin{enumerate}
\item The time period was 71 days between August 11, 1987 and  October 20, 1987.
\item Angular difference from SN1987A was less than $10.0^{\circ}$.
\end{enumerate}
Four events were selected according to these criteria, as shown
in Table~\ref{tab:eventlist}.
The celestial map of all upward-going muon events recorded in
this time period is shown in Figure~\ref{fig:skymap} together with
the $10.0^{\circ}$ angular window around SN1987A.

The probability that these four events can be explained
by atmospheric neutrinos was calculated as follows.
Table~\ref{tab:probability} lists the number of candidates and
the number of expected
atmospheric neutrino backgrounds from the time/angular window.
Atmospheric neutrino backgrounds were estimated
based on the publications from
the experiments \citep{SN1987Aupmu,Kamupmu,Hara,IMBupmu}. 
When atmospheric neutrino background from
SN1987A was not specified in the papers, the background for
LMC~X-4 (RA=83.2$^{\circ}$, $\delta$=$-$66.4$^{\circ}$), whose celestial
position is only 2.9$^{\circ}$ away from SN1987A, was used.
For the upward stopping muon analysis in Kamiokande-II,
the zenith angle averaged stopping/through muon ratio, 0.20, was used \citep{Hara}.
The probability that the number of events is larger or equal
to true observation was also calculated.
The details of the calculation and the results are listed
in Table~\ref{tab:probability}.

Based on these individual probabilities,
the probability ($P$) for the chance coincidence to detect 4 atmospheric neutrino
events in the assumed time/angular window was calculated as:
$$
P = 0.119 \times 0.026 \times 0.040 \times (385/71) \times 4 = 0.0027.
$$
The factors $385/71$ and 4 are \char'134 trials factors".
The factor $385/71$ is based on the time window; a 71 day window was employed
over the 385 days of data that were analyzed.
The factor 4 means that, among four event categories (Kam-thru, Kam-stop,
IMB-thru and IMB-stop), three event categories except IMB-stop were
selected and used in the calculation.
Accordingly,
$$
_{4}{\rm C}_{3} = {{4!}\over{3!(4-3)!}} =4
$$
was multiplied by the probability.
The probability that these signals are a chance coincidence of atmospheric
neutrinos was found to be 0.27\%.

It should be noted that the probability for chance coincidence
for other time/angular windows can be easily calculated by
the similar procedure.
The expected atmospheric neutrino background ($\lambda'$) for the time duration
$\Delta T$ days and angular radius $\Delta\theta$ degree can be obtained by
a scaling of the numbers in Table~\ref{tab:probability}:
$$
\lambda' = \lambda \times {{\Delta T}\over{71}}
        \times {{1-\cos(\pi\Delta\theta/180^{\circ})}\over{1-\cos(\pi 10^{\circ}/180^{\circ})}}.
$$

\begin{figure}[!b] 
\centering
\includegraphics[width=20pc]{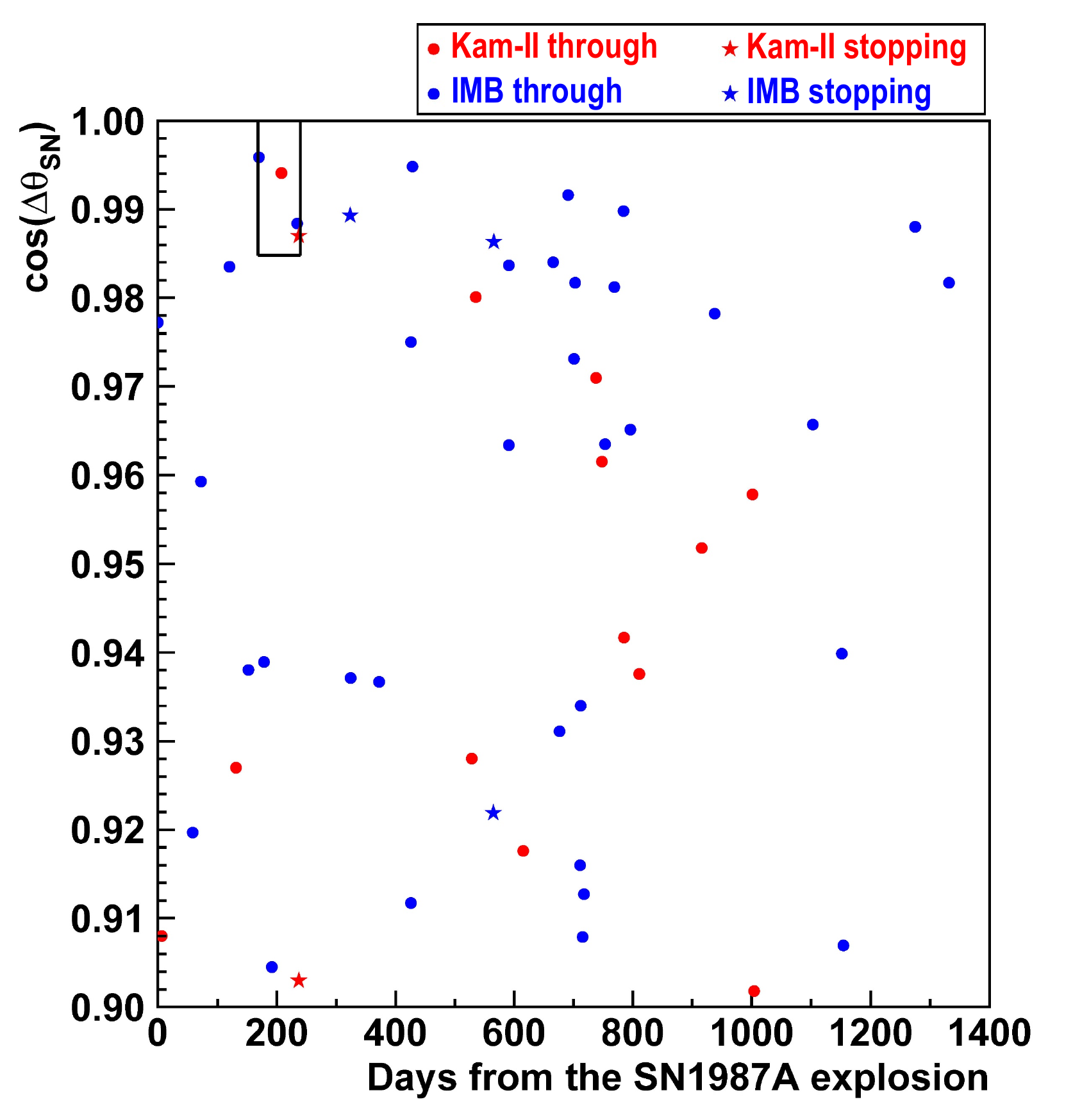}
\caption{
Angular correlation between upward-going muons and
SN1987A between the SN1987A explosion and the end of experiments.
The last upward-going muon events were recorded on April 8, 1990
(Kamiokande-II) and December 14, 1990 (IMB). They correspond to
the 1140th day and the 1390th day after the explosion, respectively.
The time/angular window between August 11 and October 20, 1987,
and $10.0^{\circ}$ from SN1987A is also shown.
}
\label{fig:sn1400days}
\end{figure}

The expected time period of the high-energy neutrinos
from SN1987A is within one year from the explosion
at most \citep{Gaisser,Sato1,Sato2}.
Therefore, the analysis using data until the 385th
day reasonably covers the expected neutrino period. However, for the completeness of the analysis,
all data after SN1987A was also examined.
Figure~\ref{fig:sn1400days} shows an angular correlation between upward-going muons and
SN1987A between the SN1987A explosion and the end of experiments. 
Obviously, no event cluster from the direction of SN1987A can
be found after one year from the explosion.

To examine the probability of the chance coincidence more carefully,
a \char'134 fake event" analysis was employed.
Most of the upward-going muons are produced from atmospheric neutrinos
generated at the opposite side of the Earth.
Their times and directions are denoted by $(t_{i},(\Theta,\Phi)_{i})$,
where $t_{i}$ is the time of the event, and $(\Theta,\Phi)_{i}$ is the
zenith angle and azimuth angle of the event in the horizontal
coordinate.
From 252 upward through going muons by Kamiokande-II and
666 upward through going muons by IMB, 
fake event sets were produced by shuffling the combination of
the time $t_{i}$ and the direction $(\Theta,\Phi)_{i}$ using random numbers.
For upward stopping muons, the shuffling method was not used because
of small number of real events. Instead, fake event sets were produced based
on the zenith angle distribution of the upward-going muon flux.
A hundred thousand sets of fake upward-going muon events were produced,
and the similar analysis as for the real data were applied.
Among 100,000 event sets, only 243 event sets showed 
similar event cluster between the SN1987A explosion
and the 385th day. This result well agrees with the probability of the
chance coincidence 0.27\% obtained for the real data.

It could be argued that the angular window $10^{\circ}$ is too
large for an analysis of directional coincidence;
a $7.0^{\circ}$ window was employed
in the previous analysis \citep{Oyama}.
However, if the nominal neutrino energy is much lower than
the theoretical expectations, around $\sim$10~GeV for example,
the scattering angle by the neutrino interactions can be larger.
The angular correlation between parent neutrinos and daughter
muons in the $<$~100~GeV energy range can be written \citep{Berezenskii} as
$$
\Delta\theta_{\nu\mu} \sim 2.6^{\circ}\sqrt{(100{\rm GeV}/E_{\nu})}.
$$
For $E_{\nu}=10~{\rm GeV}$, $\Delta\theta_{\nu\mu} \sim 8.2^{\circ}$.
Noted that the angular deviation due to multiple scattering
of daughter muons during their travel and the angular resolutions of the
detectors should be also considered.
The existence of one upward stopping muon candidate
supports this ``low energy" hypothesis
because $\sim$10~GeV neutrinos can be detected as upward stopping muons
with high probability.
The ``low energy" hypothesis implicitly suggests that the nominal proton energy
accelerated by SN1987A might be $\sim$100~GeV.

\section{Summary}
In summary, 2 upward-going muons by Kamiokande-II
and 2 upward-going muons by IMB are observed between August 11 and October
20, 1987 within $10.0^{\circ}$ angular window around SN1987A. 
The probability that these events can be explained by
a chance coincidence of atmospheric neutrinos was
calculated as 0.27\%.
These events might be the first evidence
of high-energy neutrinos from a supernova explosion.
\medskip

\acknowledgments{
\section*{Acknowledgments}
The author proposed to publish this result
as a joint publication from the Kamiokande-II collaboration
and the IMB collaboration. However, remaining members of
both collaborations suggested that very few members
remain active and making a decision as collaboration is
difficult. It was also suggested that the old data are free for use, and
the results can be published with an author's responsibility.
The author gratefully acknowledges all members of both collaborations.
}

\end{document}